\pdfoutput=1
\documentclass[10pt,a4paper]{article}
\RequirePackage{latexsym,amsmath,amssymb}
\RequirePackage[dvipsnames,usenames]{color}

\usepackage{fullpage}
\usepackage{cancel}
\usepackage[british]{babel}
\usepackage[latin1]{inputenc}
\usepackage[T1]{fontenc}
\usepackage[final]{showkeys} 
\usepackage{hyperref}
\hypersetup{pdfauthor={M. Cadoni, A. Sanna...},
            pdftitle={}
            }
\usepackage[]{cleveref}
\Crefname{equation}{Eq.~}{Eqs.}
\Crefname{figure}{Fig.}{Figs.}
\crefformat{plural}{#2eqs.~(#1)#3}
\crefname{section}{Sect.}{Sects.}

\usepackage{amsfonts}
\usepackage{graphicx}
\def\lb{\label}
\def\be#1\ee{\begin{align}#1\end{align}}

\newcommand{\beq}{\begin{equation}}
\newcommand{\eeq}{\end{equation}}

\renewcommand{\ge}{\geqslant}

\renewcommand{\le}{\leqslant}

\def\a{\alpha}

\def\g{\gamma}
\def\l{\lambda}
\def\lb{\label}

\def\b{\beta}

\title{\bf How to  reduce epidemic peaks  keeping under control the time-span of the epidemic  }
\author{Mariano Cadoni${}^{ab}$\thanks{E-mail: mariano.cadoni@ca.infn.it},\
\\
\\
${}^a$\emph{Dipartimento di Fisica, Universit\`a di Cagliari}
\\
{\em Cittadella Universitaria, 09042 Monserrato, Italy}
\\
\\
${}^b$\emph{I.N.F.N, Sezione di Cagliari, Cittadella Universitaria, 09042 Monserrato, Italy}
\\
\\}
\begin{document}
\maketitle
\begin{abstract}
One of the main  challenges of the measures  against the COVID-19 epidemic is to reduce the amplitude   of the epidemic peak without increasing without control its timescale. We investigate this problem using the  SIR model  for the epidemic dynamics, for which   reduction of  the epidemic peak $I_P$  can be achieved only at the price of increasing  the time $t_P$ of its occurrence and its entire  time-span $t_E$. By means of a time reparametrization we linearize  the equations for the SIR dynamics.  This allows us to solve exactly the dynamics in the time domain  and to derive the scaling behaviour of the size,  the timescale  and the speed of the epidemics, by reducing  the infection rate $\alpha$  and  by increasing  the removal rate $\beta$ by a factor of $\lambda$.   We show  that for  a given  value of  the size ($I_P$, the total, $I_E$ and average $\hat I_P$  number of infected),   its occurrence  time $t_P$ and entire time-span $t_E$ can be reduced by a factor $1/\lambda$ if the reduction of $I$ is achieved by increasing the removal rate instead of reducing the infection rate. Thus, epidemic containment measures based  on tracing, early detection followed  by prompt isolation  of infected individuals are more efficient than those based on social distancing.   We apply our results to the  COVID-19 epidemic in Northern Italy. We show that the  peak time $t_P$  and the entire time span $t_E$ could have been   reduced by  a factor  $0.9 \le 1/\lambda\le 0.34$ with  containment   measures  focused on increasing $\beta$ instead of  reducing $\alpha$.  
\end{abstract}
\section{Introduction}
The recent COVID-19 epidemics is posing   formidable challenges  both to the health and  economic systems  worldwide. 
In order to tackle  the ongoing epidemic, the  countries faced with the COVID-19 epidemic  have used different strategies.  In order   to control the epidemic one has to lower the basic reproduction number (also called basic reproduction rate) $\rho$, eventually   below the threshold, i.e  $\rho<1$. In a simple SIR Model \cite{Kermack,Murray,Hethcote,Brauer,Daley,Brauer1}, which we consider in this paper,  $\rho$ depends on the infection  rate $\alpha$, the removal rate $\beta$ and the total population $N$. If $\rho>1$ the epidemic starts with  the number of infected individuals  $I$ growing exponentially,  it  reaches a maximum  $I_P$   at a time $t_P$, then decreases   down to zero at time $t_E$ when the epidemic ends. 

Although for severe epidemics like the COVID-19 it is very difficult to keep the reproduction number $\rho$ from the beginning  below the  threshold, nevertheless  containment measures  soften the epidemic because they  lead to   a reduction of the epidemic peak $I_P$.

 Basically, there are therefore  three  types of containment  measures  one can use.   	One can  act on $\alpha$ by lowering it,  for instance  by forcing social distancing or by prophylaxis measures.  One can achieve the same result   increasing $\beta$, e.g. by means of a prompt strict isolation  of infected individuals.  Last but not least  one can reduce the total population $N$ by separating it in strictly non-communicating compartments. 

The policies  of different countries  for  fighting  the COVID-19 usually contemplate a mixture  of the three types of measures mentioned above. Obviously,  the choice of focusing on one kind  of measure instead of another depends on  a number of  factors, which include not only its  effectiveness  but also its  feasibility and its  social and economic impact.  For instance most of the European countries have focused their COVID-19 fighting strategies on measures  aimed at  reducing the  infection rate  $\alpha$. Countries like Korea and Singapore favoured instead  measures aimed at rising $\beta$.

A general, unpleasant,   feature of the  epidemic  dynamics is that the reduction of the epidemic peak $I_P$ can be only achieved  at the expenses  of increasing $t_P$ and $t_E$, i.e of increasing the time-span of the epidemic. 
On the one hand for  epidemics like  COVID-19 having a high rate of  hospitalised  infected with  severe symptoms,  reduction of $I_P$ and dilution of the epidemic  is necessary in order  to allow the health systems to  treat them properly.  On the other hand,  increasing $t_P$ means  increasing the time-span in which the containment measures are  effective,  with  potentially disruptive   effects both on the economies and the  live   of the populations in the involved countries.   

In this paper we investigate, in the framework of  the SIR model,  the impact of containment measures,  which act on  $\alpha$  and $\beta$, on $I_P$,  $t_P$, $t_E$  and on the epidemic speed
($d\rho/dt|_P$). The structure of the paper is as follows.
In Sect \ref{sec:2}, by means of   a time reparametrization, we are able to linearize  the equations for the SIR dynamics. This allows us to solve exactly the dynamics in the time domain  and to derive the scaling behaviour of  $I_P$,  $t_P$, $t_E$ and  $d\rho/dt|_P$, by  reducing  the infection rate $\alpha\to\alpha/\lambda$  or by increasing the removal rate $ \to \lambda \beta$.   We show  that keeping  $I_P$ fixed  its occurrence  time $t_P$ and $t_E$ can be  be reduced by a factor $1/\lambda$ by acting  on the removal rate $\beta$ instead on the infection rate  $\a$. This  will be discussed in  Sect. \ref{sec:3}. In Sect. \ref{sec:4} we discuss approximate solutions   of the SIR model in which the reproduction number $\rho<e$.
In Sect. \ref{sec:5} we apply our results to the  COVID-19 epidemic in Northern Italy. We show that the  peak time $t_P$  and the entire time-span of the epidemic could have been   reduced by  a factor  $0.9 \le 1/\lambda\le 0.34$ with  containment   measures  focused on increasing $\beta$ instead of  the reducing $\alpha$.  Finally, in Sect. \ref{sec:6} we state our conclusions.

\section{SIR model: time reparametrization  and linearization} 
\lb{sec:2}
The SIR model describes the deterministic dynamics of an infective epidemic, characterized by the fact that individuals, which have been infected and have recovered  gain permanent immunity \cite{Kermack,Murray,Hethcote,Brauer,Daley,Brauer1}.
Although the model is quite simple, it can be used to give at least  rough estimates of  epidemic dynamics, and in particular of the COVID-19 epidemic \cite{Gaeta1,Gaeta3,Gaeta4,Gaeta5}.  A generalisation of the SIR model to take into account a large number of asymptomatic infectives -hence more apt to  describe the COVID-19 epidemic- has been proposed in Ref. \cite{Gaeta3,Gaeta4,Gaeta5}.    

The homogeneous and isolated population of $N$ individuals exposed  to the epidemic,   is characterised at time $t$ by the number of susceptible  $S(t)$, infected and infectives $I(t)$  and removed (recovered, dead or isolated) $R(t)$ individuals, with  the conservation  law  $N=S(t)+I(t)+R(t)$. The timescale  of the epidemics  is assumed to be relatively short so that $N$ can be assumed   constant.

The dynamic describing the evolution of the epidemic is deterministic and described by the following,  non linear,  dynamical system:
\begin{subequations}
\begin{align}
&\frac{dS}{dt}=-\alpha SI,\label{sir1}\\
&\frac{dI}{dt}=\alpha SI-\beta I,\label{sir2}\\
&\frac{dR}{dt}=\beta I.  \label{sir3}
\end{align}
\end{subequations}

The infective epidemic is characterised by two parameters: 
$(1)$ The infection rate  (also called contact rate) $\alpha$, which gives the transition rate between the class of susceptible and  that of infected;  
 $(2)$ the removal rate  $\beta$, which gives the transition rate between the class of infected  and  that of removed  ($1/\beta$ gives the characteristic time for the removal of infected from the dynamics).

 From equation (\ref{sir2}) it is immediately evident that number of infected individuals  grows, i.e the epidemic spreads, only if 
\beq
\lb{soglia}
S>\gamma, \quad\quad \gamma:=\frac{\beta}{\alpha},
\eeq  
where  $\gamma$ is the epidemic threshold.  
Equivalently, one can introduce the   basic reproduction number $\rho(t)$:
\beq
\lb{cr}
\rho(t)= \frac{S(t)}{\gamma},
\eeq 
which represents  the expected number of new infections generated by  a single infection. The epidemic spreads if $\rho>1$. The parameters $\a$ and $\b$ depend on   several factors.  Some of them are attributes of the pathogen causing the disease and cannot be changed. Other are influenced by the social behaviour of the individuals and can be therefore changed  with containment and prophylaxis measures. 

The system (\ref{sir1}),(\ref{sir2}),(\ref{sir3})  is difficult to solve,  analytically in the time domain. Usually one proceeds by eliminating  $dt$ from  Eqs. (\ref{sir1}),(\ref{sir2}), then after integration one easily finds the function $I(S)=I_0+(S_0-S)+ \g \log(S/S_0)$, where  $I_0,S_0$ are the initial data  (see e.g. Ref. \cite{Gaeta4}). This form of $I(S)$ allows one to derive some qualitative and quantitative  features of the epidemic dynamics but not its explicit time evolution. This latter can be only obtained  by numerical integration of Eqs. (\ref{sir1}),(\ref{sir2}),(\ref{sir3}). For instance, one can easily find that, if initially we are above  the threshold $\rho_0>1$, $I(t)$ grows till it reaches a maximum $I_P$, then it goes down to zero at a time $t_E$ when the epidemic ends. 

The function $I(S)$ allows to determine analytically the value of the peak  $I_P$  but not the  time $t_P$ of its occurrence, nor its entire time-span $t_E$, nor the speed    of  the reproduction number $V_P:=d\rho/dt|_P$, nor the average  value of the number of infected individuals at the peak $\hat I_P$. $t_P$, $t_E$, $V_P$ and $\hat I_P$  have to be determined  after solving numerically the dynamics. This is a quite unpleasant  feature because it prevents a clear understanding of the dependence  of $t_P$,$t_E$, $V_P$ and $\hat I_P$ from the parameters $\a,\b$,  which  is a crucial information for fighting the epidemic.

In order to solve analytically the temporal dynamics let us reparametrize  the time   introducing  a new time coordinate $\tau$ defined by $d\tau/dt= I(\tau)$, i.e.: 
\beq\lb{tau}
t-t_0=\int_{\tau_0}^\tau \frac{d \tau'}{I( \tau')},
\eeq
where $t_0=t(\tau_0)$ is the initial time.   Using   time-translations we can put without loss of generality,  $\tau_0=t_0=0$.  The new time coordinate  has a simple intuitive meaning,  $\tau(t)/t$  gives the average value $\hat I(t)$ the number of infected at time $t$:
\beq\lb{tm}
\hat I(t):= \frac{1}{t}\int_{0}^t  I(t')dt'= \frac{\tau(t)}{t}.
\eeq
The time reparametrization (\ref{tau}) allows to linearise the system (\ref{sir1}),(\ref{sir2}),(\ref{sir3}):
 \begin{subequations}
\begin{align}
&\frac{dS}{d\tau}=-\alpha S,\label{sirt1}\\
&\frac{dI}{d\tau}=\alpha (S-\g) ,\label{sirt2}\\
&\frac{dR}{d\tau}=\beta .  \label{sirt3}
\end{align}
\end{subequations}
This  can be easily integrate to give: 
\begin{subequations}
\begin{align}
&S=S_0e^{-\a \tau},\label{sol1}\\
&I=I_0+S_0-S_0e^{-\a \tau}-\beta \tau ,\label{sol2}\\
&R=R_0+\beta\tau ,  \label{sol3}
\end{align}
\end{subequations}
where $S_0,I_0$ are initial data and $R_0= N-S_0-I_0$. In the following we will take $R_0=0$.
The function $\tau(t)$ is defined implicitly by
\beq\lb{ttau}
t=\int_{0}^\tau \frac{d \tau'}{I_0+S_0-S_0e^{-\a \tau'}-\beta \tau'}.
\eeq
Exact solutions of  the  SIR model, which are  equivalent  to our  Eqs. (\ref{sol1}),(\ref{sol2}),(\ref{sol3}), (\ref{ttau}) have been derived in Refs.  \cite{Harko,Miller} using  a  completely different approach.

Although, the integral (\ref{ttau}) cannot be evaluated analytically,  it allows to solve  the temporal dynamics of the SIR model and  to investigate the scaling behaviour of the  relevant quantities characterizing the epidemic when  the parameters $\a,\b$ change. 
The previous expressions allow us  to compute easily all the relevant quantities  for the epidemic  peak.  From Eqs. (\ref{sirt2}) and (\ref{sol1}) one gets immediately  $\tau_P$, the $\tau$-time coordinate of the peak. The other quantities are readily  computed using Eqs.  (\ref{sol1}),\ref{sol2}),(\ref{sol3}), (\ref{ttau}): 
\begin{subequations}
\begin{align}
&I_P=I_0+S_0- \g -\g \log\left(\frac{S_0}{\g}\right),\label{peak1}\\
&S_P=\g,\quad  R_P= \g\log\left(\frac{S_0}{\g}\right),\quad  V_P=-\a I_P,\label{peak2}\\
&\tau_P=\frac{1}{\a}\log\left(\frac{S_0}{\g}\right),  \label{peak3}\\
&t_P=\int_{0}^{\tau_P} \frac{d \tau'}{I_0+S_0-S_0e^{-\a \tau'}-\beta \tau'}.\label{peak4}
\end{align}
\end{subequations}

The entire time-span of the epidemic $t_E$ can be computed setting $I=0$ in Eq. (\ref{sol2}). Because $I_0$ is usually small compared to $S_0$ it can be  neglected, $t_E$  is obtained by first finding the (higher)  root  $\tau_E$ of the transcendental equation:  
\beq\lb{ete}
S_0-S_0e^{-\a \tau_E}-\beta \tau_E=0,
\eeq
and then using Eq. (\ref{ttau}) to compute $t_E$.

An other important quantity, which describes the intensity of the epidemics is the total number $I_E$ of individuals that are infected over the whole time-span of the epidemics.  Taking into account that $I_0$ is rather small and that initially $S_0=N$, $I_E$ can  expressed in terms of the lower root $S_E$ of the transcendental equation obtained by setting $I=0$ in Eq. (\ref{sol2}) (see Ref. \cite{Gaeta3}) for details. We have 
\beq\lb{it}
 I_E=N-S_E,
\eeq
where $S_E$ is the lower root of the transcendental equation 
\beq\lb{eit}
N-S_E+\g \log\left( \frac{S_E}{N}\right)=0.
\eeq
\section{Scaling behaviour of epidemic parameters}
\lb{sec:3}
In this section we investigate the scaling behavior of the peak quantities  (\ref{peak1}...(\ref{peak4}), the total number of infected (\ref{it}),  $I_E$ and $t_E$  by  changing of the parameters $\a$ and $\b$.  It is  already known  that Eqs. (\ref{sir1})...(\ref{sir3}) are invariant under the scaling $\a\to \l\a, \, \b\to \l \b,\, t\to \l^{-1} t$ \cite{Gaeta3}. This scaling transformation    leaves invariant the epidemic threshold  $\gamma$ and tells us that we can   increase  (reduce)   the timescale of the epidemic by {\sl simultaneously}   reducing (increasing)  both $\a$ and $\b$.  However, this is not what we are interested in. Actually,  we want to  know  what happens  to the  epidemic parameters listed above   when we increase the threshold  $\g$.
Let us first observe that  both the number of infected at the peak $I_P$ (see Eq. (\ref{peak1})  and the total number of infected $I_E$ (see Eq. (\ref{it})  are decreasing functions of the parameter $\g$. In fact, we get from Eq. (\ref{peak1}) and Eq. (\ref{eit}),
\beq\lb{dg}
\frac{dI_P}{d\g}= - \log\left(\frac{S_0}{\g}\right),\quad \frac{dS_E}{d\g}= - \left(\frac{\g}{S_E}+1\right)^{-1}\log\left(\frac{S_E}{N}\right).
\eeq
We see that above the epidemic threshold ($S_0/\g>1$),  $dI_P/d\g$ is always negative, while  being $S_E<N$, $dS_E/d\g$ is always positive.

It  follows  that if  we want to reduce the peak  and the total number of infected  we  have to increase $\gamma$ by a factor  $\l>1$.
Because  one can increase $\gamma$ either by {\sl reducing} $\a$ or by {\sl increasing} $\b$, we have to   compare the effects on the peak parameters of these two different ways of  increasing  $\g$. 

We are therefore lead to consider two different  scaling transformations:
transformation $T^{(1)}$, which reduces  the infection rate: 
\beq\lb{st1}
 \a\to \l^{-1} \a,\quad \g\to \l \g,\quad  \l\ge 1
 \eeq 
and transformation $T^{(2)}$, which increases   the removal rate:
\beq\lb{st2}
 \b\to \l \b,\quad \g\to \l \g,\quad  \l\ge 1.
 \eeq 
 
The peak quantities in Eqs. (\ref{peak1}...(\ref{peak4}) and the $\tau_E$  of Eq. (\ref{ete}) do not transform in a simple way under $T^{(1)}$ and $T^{(2)}$, however the ratios of  $T^{(1)}$ and $T^{(2)}$-transformed  quantities follow simple scaling laws. In particular, they remain invariant whenever the quantity depends only on their ratio  $\g$ and not on $\a$ and $\b$ separately.

Using the following 
notation to denote  rescaled quantities: $I^{(1)}_P=I_P(\l^{-1}\a),\quad I^{(2)}_P=I_P(\l\b)$ and similarly for the others quantities, we get,
\begin{subequations}
\begin{align}
&I^{(1)}_P=I^{(2)}_P,\quad\quad S^{(1)}_P=S^{(2)}_P,\quad \quad R^{(1)}_P=R^{(2)}_P,\label{resc1}\\
&V^{(2)}_P=\l V^{(1)}_P,\quad \quad \tau^{(2)}_P=\l^{-1}\tau^{(1)}_P, \quad\quad \tau^{(2)}_E=\l^{-1}\tau^{(1)}_E.\label{resc2}
\end{align}
\end{subequations}
The transformation law for $t_P$ and $t_E$ can be derived  by  first acting with the transformation  $T^{(1)}$ on the integral (\ref{peak4}), then acting with $T^{(2)}$ on the same integral  and finally  redefining  the integration 
variable in the second integral $\tau'\to \l^{-1} \tau'$. One obtains in this way:
\beq\lb{stP}
t^{(2)}_P=\l^{-1}t^{(1)}_P,\quad \quad t^{(2)}_E=\l^{-1}t^{(1)}_E.
\eeq
Finally, using Eqs. (\ref{tm}), (\ref{resc2}), (\ref{stP})  and taking into account that $I_E$ depends only on $\g$ (see  Eqs. (\ref{it}),(\ref{eit})) we can easily show that the average  number of infected    
$\hat I_P$ and  the total number of infected $I_E$ are  invariant, i.e.
\beq\lb{stPM}
\hat I^{(2)}_P=\hat I^{(1)}_P, \quad\quad I_E^{(1)}=I_E^{(2)}.
\eeq
An important result follows from equations (\ref{resc1}),(\ref{resc2}),(\ref{stP}) and (\ref{stPM}): epidemic containment measures, which have the same effect for what concerns  $I_P$,  $R_P$,   $S_P$, $\hat I_P$ and $I_E$,  have  different impact on the  occurrence time  $t_P$ of the peak, the whole time span of the epidemic $t_E$ and on the epidemic speed $V_P$. Choosing measures  increasing the removal rate $\b$ by a factor $\l$ instead of reducing the infection rate $\a$  by a factor $1/\l$ allows to drop $t_P$ and $t_E$ by a factor $1/\l$. For instance  by implementing epidemic containment measures  with   $\l=2$ we can  reduce by a half both  the time needed for the epidemic to reach the peak and the whole time-span of the epidemic.  It should be noticed that this epidemic timescale   reduction effect becomes more relevant for epidemics with high reproduction number  $\rho_0>>1$. In fact the  factor $\l$ is limited by $\l<\rho_0$, simply because for $\l>\rho_0$ the epidemic does not develop at all. Thus,  if we have for instance $\rho_0=5$ we can reduce the  peak time and the entire time-span of the epidemic until  a factor of $1/5$. 

Therefore, increasing $\b$  represents an efficient  way to  fight epidemics.  If  by increasing it we manage to bring $\rho$ below the threshold we simply stop the epidemic, but even if we do not go so far, we can still  reduce the size of an epidemic keeping under control its timescale.

The behaviour of the  reproduction number  speed $V_P$ in Eq. (\ref{resc2}) explains  clearly what is going on. If one acts on $\b$  instead on $\a$,  $V_P$   increases, as expected,  by a factor of $\l$.  In short, increasing   $\b$  instead of reducing $\a$, allows one to speed up the epidemic dynamics keeping constant   the number of infected at the peak, the average number of infected and the total number of infected. This is possible because the increasing of the removal rate allows prompt removal of infected individuals.

\section{ Approximate solutions   for $\rho_0<e$}
\lb{sec:4}
In the general case the integral (\ref{ttau}) cannot be computed analytically. Therefore the function $\tau=\tau(t)$ has to be computed numerically, by first  performing numerical integration of the integral in (\ref{ttau}) to find $t=t(\tau)$ and then inverting it.  There is, however, a situation  in which the integral (\ref{ttau}) can be computed analytically and the dynamics  of the epidemic until the peak, can be expressed analytically in closed form in terms of the time $t$, albeit in approximate  form. 

For   $\a\tau<<1$ we can approximate the exponential in Eq. (\ref{ttau}) by $e^{-{\a\tau}}\approx 1-\a\tau$. This approximation allows to solve the integral and to invert the function $t=t(\tau)$. We find,
\beq\lb{aa}
\tau \approx \frac{I_0}{\a(S_0-\g)}\left(e^{\a(S_0-\g)t}-1\right).
\eeq
With this position we can easily  write down the approximate form of  the solutions (\ref{sol1}),(\ref{sol2}),(\ref{sol3})  in terms  of the time  $t$. We quote here only the form  of  $I(t)$ and $t_P$ as a function of  $I_P$,
\beq\lb{asol}
I(t)\approx I_0 e^{\a(S_0-\g)t},\quad   t_P\approx \frac{1}{\a(S_0-\g)}\log\frac{I_P}{I_0}.
\eeq
Eqs.   (\ref{aa}) and (\ref{asol}) are a good approximation only for $\a\tau<1$. Because $\tau(t)$,  is an increasing function of $t$, the approximation for the dynamics   is good until the  peak, if $\a\tau_P<1$, which implies from equation (\ref{peak3}):
\beq\lb{ap}
\rho_0=\frac{S_0}{\g}<e.
\eeq

\section {Application to the COVID-19 epidemic in Northern Italy}
\lb{sec:5}
The recent development  of the COVID-19 epidemic in Northern Italy represents an interesting case for  applying    the results described in the previous sections.  The epidemic developed in the three main regions  of Northern Italy (Lombardia, Veneto and Emilia-Romagna) we consider in this paper,  starting from  end of February 2020 (although it may  be possible that the epidemic was circulating  in the regions before that date). 

 Altogether  these  three  regions have around 20 millions of inhabitants, we will therefore take $N=2 \cdot10^7$ in our  computations. We  take  as initial value $I_0=100$, which approximately corresponds to the  known cases of COVID-19 infected    Northern Italy on February 23, 2020.
The determination of the initial values   of the  other two parameters of the SIR model $\a_0$ and $\b_0$ is   more involved.  These initial values are completely determined by the pathogen because they are not affected by the epidemic containment measures put into play. The value of $\a_0$ can be determined  from the  exponential  behaviour of the early dynamics, or equivalently from the initial doubling time \cite{Gaeta1,Gaeta2}. 
Using the  raw data  for the early dynamics  of the epidemic  in Northern Italy, Gaeta \cite{Gaeta1,Gaeta2}  has given the estimate:
\beq\lb{a0}
\a_0= \frac{10^{-7}}{6}.
\eeq 
The determination of $\beta_0$ is even more problematic. This is because we expect it to be  sensitive  to the presence of large cohort of asymptomatic infectives. 
We can estimate $\b_0$ from the basic reproduction number $\rho_0$, using Eq. (\ref{cr}).
Rough evaluations  of $\rho_0$ give a number between  $2$ and $2.5$, however the indeterminacy related to the presence of  large number of  asymptomatic infectives may result in  a much higher value for $\rho_0$. To be  rather conservative  we  assume here $\rho_0=3$, so that Eqs. (\ref{cr}) and (\ref{a0}) give:
\beq\lb{b0}
\b_0=\frac{1}{9}.
\eeq

The COVID-19 containment  strategies put  in place in Northern Italy  are  a mixture of social distancing, social confinement, early detection and infection tracing. Although  mainly focused  of  social distancing,   these strategies contain  all the previous ingredients, which  modify  in  different ways  the parameters $\a$ and $\b$. Social distancing  acts by reducing $\a$, whereas systematic,   prompt, and strict isolation of infected individuals   as a result of early detection and tracing  enhances $\beta$. 
Both effects  rise $\g$ and reduce in the same way the amplitude of the peak $I_P$, the average number of infected $\hat I_P$ and the total number of infected $I_E$.

Because it is almost impossible to disentangle  the effects  of the various containment  measures on $\a$ and 
$\b$ in the real situation, we will discuss and compare  two hypothetical situations in which the rising of $\gamma$, 
$\gamma\to \l \gamma_0$, with $\gamma_0=(2/3)\cdot10^7$,  is obtained  in two fully distinct and complementary ways:      
\begin{itemize}
\item
$(1)$ We have exclusively social  confinement  containment  measures: $\b$ is held fixed to its initial value  $\b_0$, whereas $\a_0$ is reduced by a factor $1/\l$.
\item
$(2)$  We have exclusively containment  measures consisting in  prompt and strict isolation of  infected individuals   triggered by early detection and tracing of infected: $\a$ is held fixed to its initial value  $\a_0$, whereas $\b_0$ is increased  by a factor $\l$.
\end{itemize}
  Being $1\le \l <\rho_0$ we consider the following values:   $\l= 1,1.5,2,2.5,2.9$.  
Using Eqs. (\ref{peak1})...(\ref{peak4}) we compute for these  values  of the parameters $\a$ and $\b$  the  peak quantities:  peak amplitude $I_P$, the time  $t_P$ (in days) of occurrence of the peak  (computed by numerical evaluation of the integral (\ref{peak4})), average number of infected $\hat I_P$ and absolute value of the epidemic speed at the peak $|V_P|$ (in days$^{-1}$).  Moreover, using Eqs.  (\ref{it}), (\ref{eit}), (\ref{ete}), together with Eq. (\ref{peak4}), we compute numerically the total number $I_E$  of infected individuals during the epidemic and its whole time span $t_E$ (in days).   The results are shown in Table I.  Our results are in accordance  with the scaling   behaviour  given by Eqs. (\ref{resc1}),(\ref{resc2}), (\ref{stP})and (\ref{stPM}).

We see from Table I  that the raising  of the epidemic  threshold for $\g$ from the initial value $\g_0$ first to $2\g_0$ then to $2.9\, \g_0$ let both the number of infected  at the peak and  their average number drastically sink from the order of magnitude  $10^6$ first to $10^5$ and then to $10^3-10^4$. This reduction is the same independently of the fact that if it is achieved by reduction of $\a$ (way $(1)$) or by increase of $\b$ (way $(2)$). Similarly, the total number $I_E$ of infected individuals drops from the huge value \footnote {Notice that without containment measures at the end of the epidemic almost  all individuals have been infected.} $1.88\cdot 10^7$ till $1.16\cdot 10^7$ (for $\l=2$) and then to $1.3\cdot 10^6$  (for $\l=2.9$)

On the other hand,  the  two ways of reducing   $I_P$ and $\hat I_P$ and $I_E$ affect differently the occurrence time of the peak $t_P$ and the whole time span of the epidemic $t_E$.  By acting on $\beta$  (way $(2)$) instead of on $\alpha$ (way $(1)$)   we can shorten these times by   $33\%$ (for $\l=1.5$), by $50\%$  (for $\l=2$) and even reduce it by almost $1/3$ (for $\l=2.9$). Correspondingly,  the speed of  variation of the basic reproduction number $|V_P|$ will be enhanced by the same factors.

If the containment measures can manage to increase $\l$ above $3$, we go below   the threshold for $\rho$ and the epidemic does not start at all. Obviously, in a real situation  reducing  $\a$ or  increasing  $\b$ is  not performed  once for all at the beginning,  but occurs   in steps. 
Our main result is that in order to try to stop the epidemic it is  much more convenient   to rise $\beta$ instead of lowering $\a$ because even if we do not manage to stop it,  we are able to reduce its size and at the same  time to shorten its timescale.

\begin{center}
\begin{tabular}{|l||c|c|c|c|c|c|c|c||}
\hline
$\l$ & $\a/\a_0$ & $\b/\b_0$& $I_P$&$\hat I_P$&$t_P$(days)&$|V_P|$(days)$^{-1}$&$I_E$&$t_E$ (days)\\
\hline
1 & $1$ & 1& $6 \cdot 10^6$ & $1.13\cdot 10^6$& 58&$10^{-1}$&$1.88\cdot 10^7$&187\\
\hline
1.5 & $0.66$ & 1& $3.07 \cdot 10^6$ & $5.74\cdot 10^5$& 109&$3\cdot10^{-2}$&$1.59\cdot 10^7$&279\\
1.5 & $1$ & 1.5& $3.07  \cdot 10^6$ & $5.74\cdot 10^5$& 72&$5\cdot10^{-2}$&$1.59\cdot 10^7$&187\\
\hline
2 & $0.5$ & 1& $1.26 \cdot 10^6$ & $2.45\cdot 10^5$& 198&$10^{-2}$&$1.16\cdot 10^7$&453\\
2 & $1$ & 2& $1.26  \cdot 10^6$ & $2.45\cdot 10^5$& 99&$2\cdot10^{-2}$&$1.16\cdot 10^7$&227\\
\hline
2.5 & $0.4$ & 1& $2.95 \cdot 10^5$ & $6.42\cdot 10^4$& 426&$2\cdot10^{-3}$&$6.2\cdot 10^6$&899\\
2.5 & $1$ & 2.5& $2.95 \cdot 10^5$ & $6.42\cdot 10^4$& 170&$5\cdot10^{-3}$&$6.2\cdot 10^6$&319\\
\hline
2.9 & $0.34$ & 1& $1.12 \cdot 10^4$ & $3.7\cdot 10^3$& 1592&$6\cdot10^{-5}$&$1.3\cdot 10^6$&3213\\
2.9 & $1$ & 2.9& $1.12 \cdot 10^4$ & $3.7\cdot 10^3$& 549&$2\cdot10^{-4}$&$1.3\cdot 10^6$&1107\\
\hline
\end{tabular}
\end{center}
\medskip

{\tt Table I.} Comparison of the effect of reduction of the infection rate $\a\to (1/\l) \a$  versus increase of the removal rate $\beta\to\l \b$  on  epidemic parameter:  peak amplitude $I_P$,  average value 
of infected $\hat I_P$,  peak time $t_P$, speed  of  basic reproduction number $|V_P|$ at the peak,  total number of infected individuals $I_E$ and whole time-span of the epidemics  $t_E$. The total population is  
$N = 2\cdot10^{7}$, $\beta_0 = 1/9$ and $\alpha_0 = (1/6) 10^{-7}$.  The values of  $I_P,\hat I_P, t_P,|V_P|,I_E,t_E$    are tabulated for  values of $\l=1,1.5,2,2.5,2.9$. For sake of clarity we also show in the table  the values of $\a$ and $\b$ corresponding to a given value of $\l$.

\section{Concluding remarks}
\lb{sec:6}
In this paper we have analysed, in the context  of  the standard SIR model for epidemic dynamics, the impact of different containment measures  on  size  (the epidemic peak $I_P$, the average number of infected $\hat I_P$ and the total number of infected $I_E$), the timescale (the occurrence time of the peak $t_P$ and the whole time-span $t_E$) and the speed (time variation of the reproduction number $|V_P|$) of epidemics. Using an exact solution for the epidemic dynamics  we have been able to  derive the scaling behaviour of these quantities under change of the two  parameters  (the infection rate $\a$ and the removal rate $\b$) of the SIR model, which can be  controlled  by the containment measures. This allowed us to compare the impact on size, timescale and speed of the  epidemic of containment measures  acting either on $\a$ or on $\b$

We have shown that for  a given reduction of  $I_P,\hat I_P, I_E$ , the  timescale and the speed  of the epidemic  are   to a great extend   sensitive   to   the kind of measures we put into play.  By increasing the removal rate  $\beta$ instead of reducing the infection rate by a factor  $\l$ one can reduce the timescale of the epidemic  by a factor $1/\l$ and increase the speed of the epidemics  by a factor $\l$. In the case we have analysed  in detail, namely the COVID-19 epidemic in Northern Italy, the  reduction   factor  $\l$, in principle     could   also  take values  around $3$.

An important point we have  not addressed in this paper is the determination of the  exact way in which the usual containment measures used to fight epidemic, impact on the values of the parameters $\a$ and $\b$. 
Whereas it is quite clear that social distancing reduces the parameter $\a$ and does not change $\b$,  the effect of   other measures like, early detection  and contacts tracing is not a priori evident. Early detection and  contact tracing    increase  $\b$ only  if implemented  on a large scale and  followed by prompt and  strict isolation of the  detected   infectives. If this is not the case, it is likely that these measures just  bring a small reduction of $\a$

The recent analysis  of Gaeta \cite{Gaeta4}  of the different strategies used  in Northern Italy  to tackle the COVID-19 epidemic seems to  confirm this result.  He found that simple  early detection   and contact tracing, while  having an impact on the epidemic peak, do not substantially affect  the timescale   of the epidemic. On  the other hand he also showed  that  contact tracing if followed by  prompt isolation  is the only efficient way  to  reduce the size  of the epidemic,  without  having to live with it a long time. The Veneto experience shows that this was one  of the  factors 
underlying the success of the  containment strategy in that region.  
Thus the main lesson one can draw from our results is that, epidemic containment measures focused   on tracing, early detection followed  by prompt removal  of infected individuals are more efficient to fight epidemics  than those based on social distancing.

Let us conclude this paper with some comments about the range of validity of our results. The SIR model is an oversimplified model for epidemic dynamics. Generalisations of it   are  necessary in order to give a good descriptions of real epidemics. For instance, in the case of the  COVID-19  epidemic a generalization of the SIR model seems to be necessary in order to take into account the presence of a large set of a asymptomatic infective \cite{Gaeta3,Gaeta4,Gaeta5}.
On the other hand,  the SIR model  gives the  bare bones of deterministic epidemic dynamics. For this reason we believe that,  at least at qualitative level, the main result of this paper - the possibility to reduce the epidemic  peak  keeping under control its  timescale  by acting on removal  rates- could remain true for generalized and improved SIR-like models.

\section*{Acknowledgements}

I thank Giuseppe Gaeta for several useful discussions and comments.

\end{document}